\documentclass{mem}
\usepackage{natbib}
\usepackage{txfonts}
\usepackage{graphicx}
\usepackage[a4paper]{hyperref}
\idline{75}{282}

\begin{document}
\def\teff{$T\rm_{eff }$}
\def\wat{H$_2$O}
\def\meth{CH$_4$}
\def\ammon{NH$_3$}
\def\logl{$\log{\rm L/L_{bol}}$}
\def\teff{$T\rm_{eff }$}
\def\teq{$T\rm_{eq }$}
\def\logg{$\log{g}$}
\def\kzz{K$\rm_{zz }$}
\def\mjup{M$\rm_{Jupiter}$}
\def\kms{$\mathrm {km s}^{-1}$}
\def\cm2s{cm$^2$~s$^{-1}$}
\def\cms2{cm~s$^{-2}$}

\title{
Cloud Formation and Dynamics in Cool Dwarf and Hot Exoplanetary Atmospheres
}


\author{
Adam J.\ Burgasser }

  \offprints{A.\ J.\ Burgasser}

\institute{
Center of Astrophysics and Space Sciences, University of California, 9500 Gilman Dr., San Diego, CA 92093, USA;
\email{aburgasser@ucsd.edu}
}

\authorrunning{Burgasser}

\titlerunning{Cool Dwarf and Hot Exoplanet Clouds and Dynamics}

\abstract{
The lowest-mass stars, brown dwarfs and extrasolar planets present challenges and opportunities for understanding dynamics and cloud formation processes in low-temperature atmospheres.  For brown dwarfs, the formation, variation and rapid depletion of photospheric clouds in L- and T-type dwarfs, and spectroscopic evidence for non-equilibrium chemistry associated with vertical mixing, all point to a fundamental role for dynamics in vertical abundance distributions and cloud/grain formation cycles. For exoplanets, azimuthal heat variations and the detection of stratospheric and exospheric layers indicate multi-layered, asymmetric atmospheres that may also be time-variable (particularly for systems with highly elliptical orbits).  Dust and clouds may also play an important role in the thermal energy balance of exoplanets through albedo effects.  For all of these cases, 3D atmosphere models are becoming an increasingly essential tool for understanding spectral and temporal properties.  In this review, I summarize the observational evidence for clouds and dynamics in cool dwarf and hot exoplanetary atmospheres, outstanding problems associated with these processes, and areas where effective synergy can be achieved.
}
\maketitle{}

\section{Introduction}

Fifteen years ago, the first examples of brown dwarfs and extrasolar planets were discovered,
sources which continue to challenge our understanding of low temperature atmospheres.  Several hundred very
low-mass stars and brown dwarfs are now known (the latter distinguished by their lack of core hydrogen fusion), encompassing two newly defined spectral classes, the L dwarfs and the T dwarfs (see review by \citealt{2005ARA&A..43..195K}).  Roughly 300 extrasolar planets have been found through Doppler shift, transit, microlensing and direct detection techniques.
Collectively, these low-temperature sources span
a broad range of mass (1~{\mjup} $\lesssim$ M $\lesssim$ 100~{\mjup}), 
photospheric temperature (100~K $\lesssim$ $T$ $\lesssim$ 2500~K),
surface gravity (3 $\lesssim$ {\logg} $\lesssim$ 5.5~cm~s$^{-2}$),
age (few Myr to few Gyr), elemental composition, rotation period (hours to days),
magnetic activity, degree of external heating and interior structure. Yet all are related by their
cool, molecule-rich, dynamic atmospheres.

While direct studies of cool dwarf atmospheres have been feasible since their discovery, it is only recently that techniques to probe exoplanet atmospheres have been realized.
The best-constrained exoplanets are those which closely orbit and transit their host stars---so-called ``Hot Jupiters''---allowing reflectance and thermal spectrophotometry 
during secondary transit (e.g., \citealt{2005ApJ...626..523C, 2005Natur.434..740D}), 
and transmission spectrophotometry during primary transit (e.g., \citealt{2002ApJ...568..377C}).  
Phase curves for non-transiting systems have also been measured (e.g., \citealt{2006Sci...314..623H}), and the recent direct detection
of planets around the young stars Fomalhaut, $\beta$ Pictoris and HR~8799 \citep{2008Sci...322.1345K,2008Sci...322.1348M, 2009A&A...493L..21L} through high contrast imaging 
techniques have opened the
door to direct investigations of exoplanet atmospheres.

Two themes are prominent in the interpretation of observational data for
low-temperature stellar, brown dwarf and exoplanetary atmospheres: clouds and dynamics.  Condensate clouds are an important source of opacity and a component in the chemical network;
they also modulate the albedos, energy budgets and atmospheric structure of exoplanets under intense irradiation.    Dynamics are responsible for the redistribution of heat and modification of
chemical abundances, and are likely fundamental to cloud formation and evolution processes.

In this review, I summarize our current observational evidence for clouds and dynamics in cool dwarfs and hot exoplanets, and note outstanding issues in their role in
emergent spectral energy distributions, long-term thermal evolution, albedo, and temporal and azimuthal variability.  
I conclude with a discussion of opportunities for synergy in future generations of cool dwarf and exoplanet models, with a view toward 3D simulations.

\section{Clouds and Dynamics in Cool Dwarf Atmospheres}

\subsection{Observational Evidence for Condensate Clouds}

\begin{figure*}[]
\centering
\resizebox{0.5\hsize}{!}{\includegraphics[clip=true]{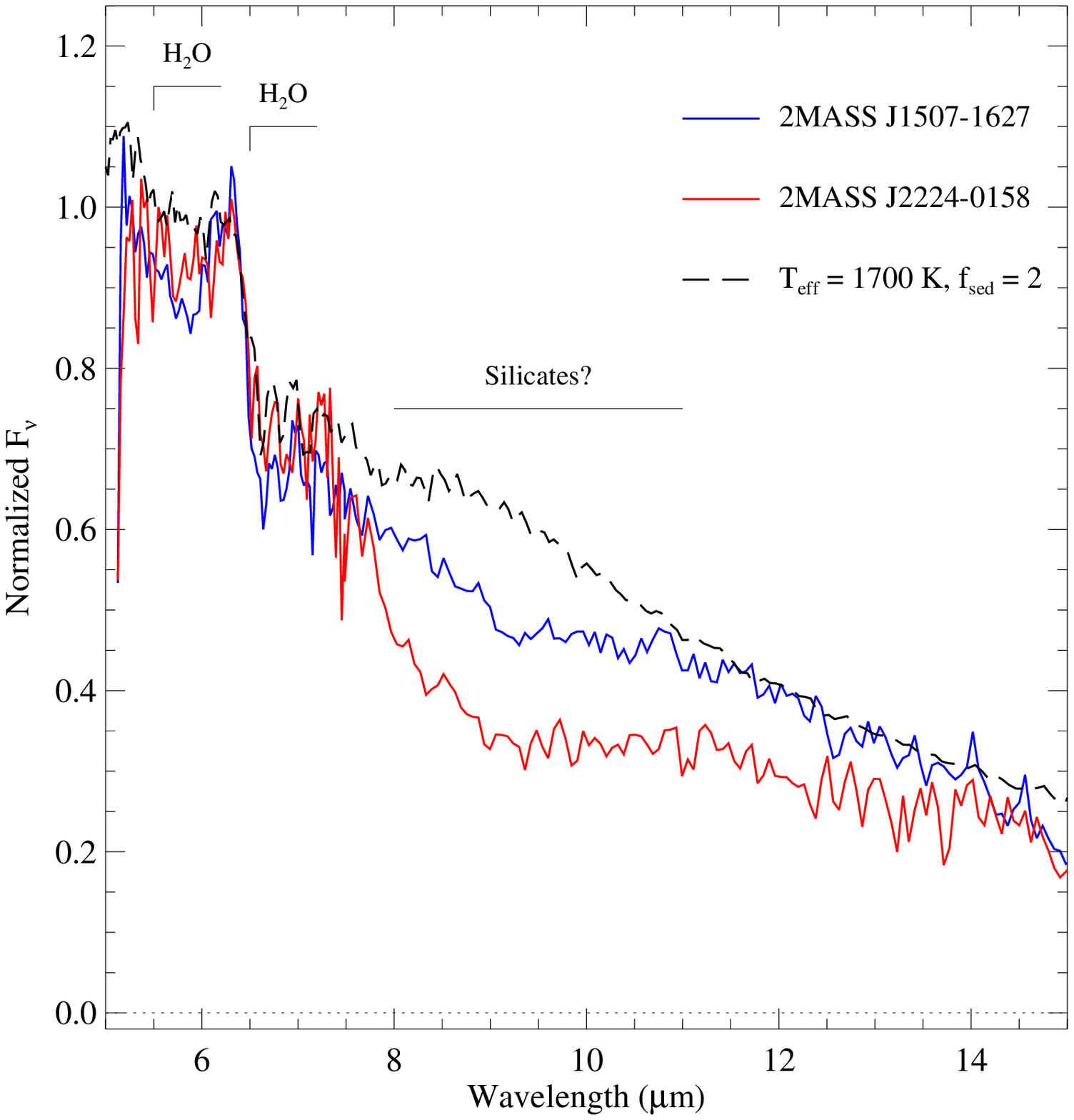}}
\resizebox{0.4\hsize}{!}{\includegraphics[clip=true]{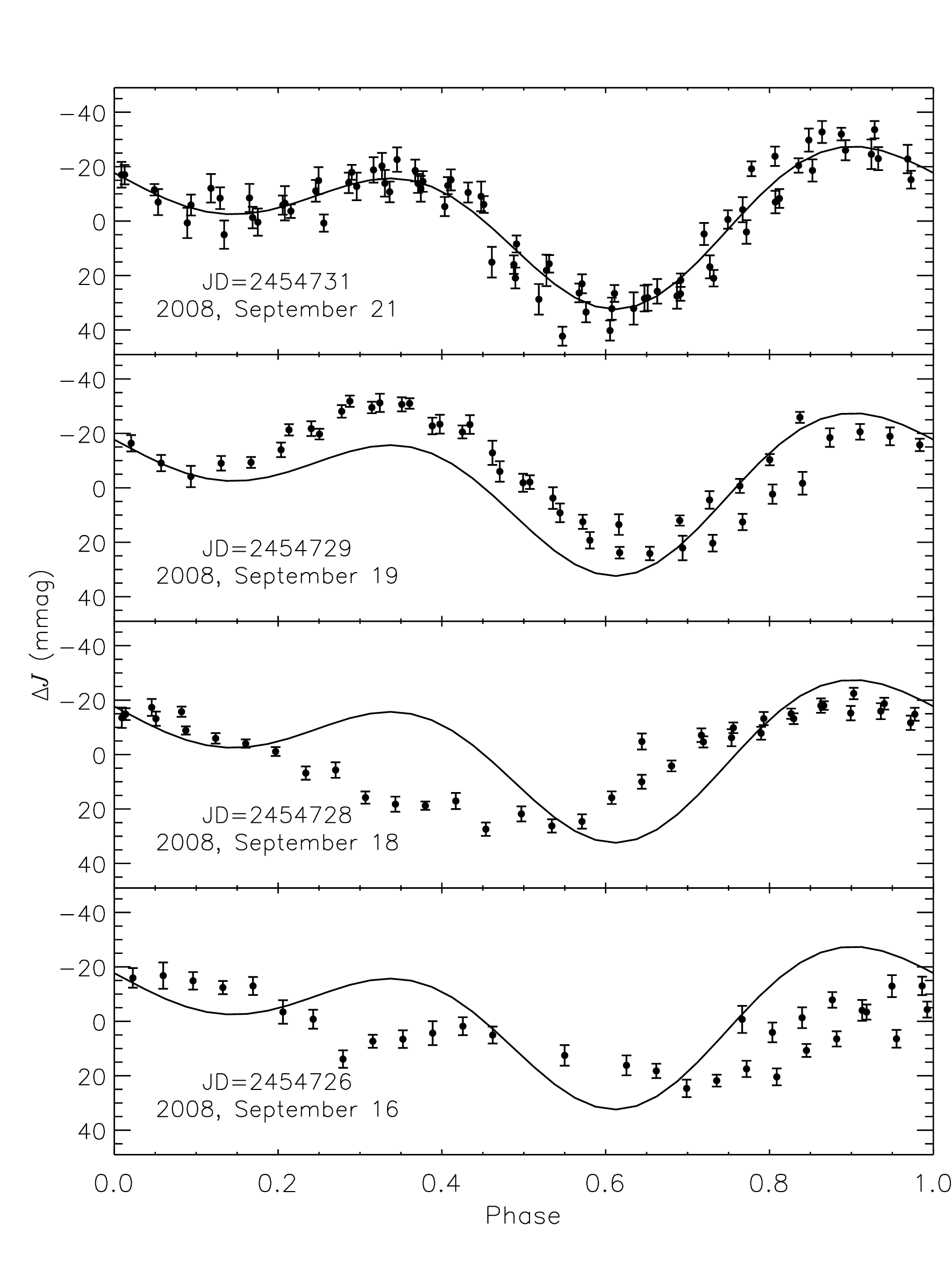}}
\caption{\footnotesize
(Left) Silicate grain absorption in the mid-infrared spectra of two equivalently-classified L dwarfs (solid lines) compared to a cloud model from M.\ Marley and collaborators (dashed line).  The L dwarf spectra also illustrate variations in the strength of this feature that correlate with near-infrared color (from \citealt{2008ApJ...674..451B}).  (Right) $J$-band light curve for the T2.5 SIMP~0136 phased to a 2.4~hr period, showing variation in both amplitude and phase over the course of 5 days (solid lines are a fit to data from 2008 September 21).  Such light curve evolution suggest a dynamic cloud layer (from \citealt{2009ApJ...701.1534A})}
\label{fig_cloud}
\end{figure*}

The atmospheres of the lowest-temperature dwarfs are
rich in molecular gas species and neutral metal elements, which produce
useful spectral discriminants for bulk properties such as effective temperature
({\teff}), surface gravity, and metallicity.  Importantly, the {\teff}s 
of L dwarfs (1400~K $\lesssim$ {\teff} $\lesssim$ 2300~K; \citealt{2004AJ....127.3516G})
span the gas/liquid and gas/solid phase transitions for
a number of refractory species, including iron, silicates, titanates, and other metal oxides
\citep{2002ApJ...577..974L}.
Evidence for these species in L dwarf photospheres has been
inferred from their very red near-infrared colors
(thermal emission from hot dust);
muted {\wat} steam bands (the result of condensate opacity at flux peaks);
and depletion of
gaseous precursors to condensates such as TiO and VO
(e.g., \citealt{1999ApJ...512..843B, 2001ApJ...556..357A}).
More recently, the {\em Spitzer Space Telescope} has directly detected
silicate grain absorption at mid-infrared wavelengths (Figure~\ref{fig_cloud}; \citealt{2006ApJ...648..614C}).
These observations make clear the prominent role of condensates in L dwarf atmospheres.

Conversely, the T dwarfs, which comprise a cooler, later stage of evolution for brown dwarfs
(600~K $\lesssim$ {\teff} $\lesssim$ 1400~K), exhibit relatively condensate-free photospheres as evident from their 
blue near-infrared colors and prominent molecular gas bands.
Strong features from neutral alkalis, whose pressure-broadened wings
absorb much of the 0.5--1.0~$\mu$m light in T dwarf spectra, also indicate that condensates have been
cleared out of the photosphere.  These species should be depleted into feldspars
in the presence of silicates; their presence requires the sequestration 
of condensates to deeper levels\footnote{A similar effect explains GeH$_4$ abundances in the atmosphere of Jupiter \citep{1994Icar..110..117F}.} \citep{2000ApJ...531..438B}.

The development of 1D condensate cloud models (e.g., \citealt{2001ApJ...556..872A})
has resolved many of these issues, although not all ($\S$~2.3).
Rather than being fully mixed in the atmosphere, clouds of condensate species are constrained between a
phase transition layer and a cloud top set by a balance of gravitational settling and mixing (the latter can be parameterized by a ``settling efficiency'' or ``cloud top temperature'').  For the L dwarfs, these cloud layers reside at the photosphere, so their influence on the spectral energy distribution is pronounced.  For the T dwarfs, the cloud layers have sunk below the photosphere, so that the emergent spectral energy distribution is largely ``condensate-free''.  It is important to note that the various condensates, while condensing at different temperatures, are probably well-mixed by convection and turbulent gas motion
(e.g., \citealt{2006ApJ...640.1063B}).  This raises the possibility of fairly complex grain chemistry, including nonequilibrium hybrid grain formation (e.g., \citealt{2008ApJ...675L.105H}).

Observational evidence supporting the presence of 
condensate clouds is also found in large
source-to-source 
near-infrared color variations among equivalently-classified L dwarfs,
at wavelengths where cloud opacity is most prominent
(e.g., \citealt{2004AJ....127.3553K}).
Early evidence suggests that near-infrared color
is correlated with the 9~$\mu$m silicate grain feature 
(Figure~\ref{fig_cloud}; \citealt{2008ApJ...674..451B}).  
In addition, temporal variability, both photometric and spectroscopic, 
has been observed in several low-temperature dwarfs
(see review by \citealt{2005AN....326.1059G}), 
most importantly those whose atmospheres are too neutral
to couple with magnetic fields and form spots (e.g., \citealt{2002ApJ...571..469M}).  
The variability likely arises from rotational modulation of 
surface asymmetries in global cloud coverage, since light curve periods
are generally consistent with rotational line broadening (e.g., \citealt{2008ApJ...684.1390R}). 
Variability amplitudes are typically $<$5\%, on par with amplitudes
in Jupiter's optical light curve (although the 5~$\mu$m variations on Jupiter are closer to 20\%; \citealt{2000ASPC..212..322G}). In addition, the rapid rotations of low-mass dwarfs ($V_{\rm rot}$ $\sim$ 30~{\kms}) and their high surface gravities ({\logg} $\sim$ 5~cgs) translate into characteristic Rhines lengths---an estimate of jet band width---and Rossby deformation radii---an estimate of vorticity scale---that are similar to Jupiter and Saturn.\footnote{The Rhines length is $(RU/2\Omega\cos{\phi})^{1/2}$, where $R$ is the radius of the body, $U$ the characteristic wind speed, $\Omega$ the rotation angular rate and $\phi$ the latitude.  Assuming $U$ $\sim$ 1~{\kms} (of order the sound speed) and $2R\Omega\cos{\phi}$ $\sim$ $V_{rot}$ $\sim$ 30~{\kms} (e.g., \citealt{2008ApJ...684.1390R}) yields a length scale of $\sim$0.2$R$.  The Rossby deformation radius is $NH/2\Omega\sin{\phi}$, where $N$ is the Brunt-V\"{a}is\"{a}l\"{a} frequency (vertical oscillation of displaced air parcels)  and $H$ the scale height.  Estimating the former as $\sqrt{g/H}$, with $g \approx 10^5$~cm~s$^{-2}$ and $H \approx 1$~km \citep{1999ApJ...519L..85G}, and $2R\Omega\sin{\phi} \sim V_{rot} \sim$ 30~{\kms}  yields a scale of  0.03$R$.  For Jupiter and Saturn, these scales are 0.14$R$ and 0.03$R$ \citep{2008ASPC..398..419S}.}  
Notably, photometric light curves are observed to evolve over time,
both in amplitude and phase, occasionally disappearing altogether (Figure~\ref{fig_cloud}).
This is an indication of surface cloud evolution over daily- to yearly-timescales.

\subsection{Observational Evidence for Dynamics}

\begin{figure}[tbp]
\resizebox{\hsize}{!}{\includegraphics[clip=true]{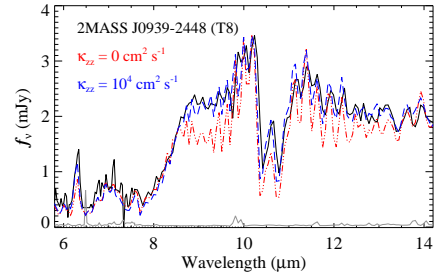}}
\caption{\footnotesize
Mid-infrared spectrum of the
late-type T dwarf 2MASS J0939-2448 (black line) compared
to {\teff} = 600~K, {\logg} = 5~{\cms2} atmosphere
models from \citet{2007ApJ...656.1136S} incorporating equilibrium chemistry 
(red dot-dashed line) and vertical mixing (blue dashed line).
Mixing is consistently required to match attenuated absorption bands of NH$_3$
in the 9--11~$\mu$m region of T dwarf spectra, 
as well as enhanced CO absorption in the 4.5--5~$\mu$m region.
}
\label{fig_ammonia}
\end{figure}

Photometric variability and variance suggest that atmospheric dynamics 
is an important contributor to cloud formation and evolution.
More direct evidence comes from the nonequilibrium  
chemical abundances of CO, {\meth} and {\ammon} observed in T dwarf spectra.
The conversion of the carbon reservoir from 
CO to {\meth} signals the transition between the L dwarf and T dwarf classes, and in chemical equilibrium this
is largely complete by {\teff} $\approx$ 1000~K (e.g., \citealt{1999ApJ...512..843B}). However, CO absorption is seen to persist well into the T dwarf regime, substantially in excess ($>$1000$\times$) of chemical equilibrium calculations
(e.g., \citealt{1997ApJ...489L..87N}).  Nitrogen chemistry---the conversion of N$_2$ to {\ammon} at {\teff} $\approx$ 700~K---is similarly in disequilibrium, with the latter molecule found to be significantly underabundant (Figure~\ref{fig_ammonia}; \citealt{2006ApJ...647..552S}).

The nonequilibrium abundances of these species can be explained by the presence of vertical mixing in the photosphere \citep{1999ApJ...519L..85G}.  Both CO and N$_2$ are strongly bonded diatomic molecules,  and their conversion to {\meth} and {\ammon} can occur over longer timescales than those vertical mixing ($\tau_{\rm mix}$).  As a result, CO and N$_2$ are found in cooler regions with abundances in excess of thermal equilibrium; equivalently, their products {\meth} and {\ammon}  are underabundant.  Current studies 
parameterize the mixing timescale with an eddy diffusion coefficient,  $\tau_{\rm mix}$ $\approx$ $H^2$/{\kzz}, where $H$ is the atmospheric scaleheight \citep{1999ApJ...519L..85G}.  Observed CO overabundances and {\ammon} underabundances can be reproduced with diffusion coefficients of order 10$^2$--10$^6$~{\cm2s} (e.g., \citep{2007ApJ...656.1136S}). These values are well below full convection, but are nevertheless indicative with vigorous mixing in the photosphere.  Importantly, mixing appears to be a universal property of L and T dwarf atmospheres \citep{2009ApJ...702..154S}.

\subsection{Dynamics and Cloud Evolution at the L Dwarf/T Dwarf Transition}

\begin{figure}[t]
\centering
\resizebox{\hsize}{!}{\includegraphics[clip=true]{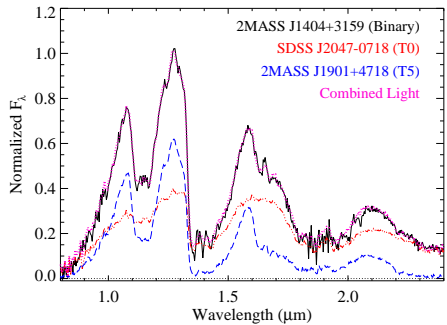}}
\caption{\footnotesize
Spectral decomposition of the resolved T0 + T5 binary 2MASS J1404-3159 \citep{2008ApJ...685.1183L}, illustrating the $J$-band bump across the L dwarf/T dwarf transition.  Spectral templates scaled to match resolved photometry and combined light spectroscopy (black line) indicate a secondary (blue dashed line) that is $\sim$50\% brighter than its primary (red dot-dashed line) at the 1.05 and 1.25~$\mu$m flux peaks.  These wavelengths are dominated by condensate cloud opacity in the L dwarfs.
}
\label{fig_binary}
\end{figure}

Current 1D cloud models have proven successful in providing a conceptual understanding of
how condensates, prominent in L dwarf photospheres, are largely absent in T dwarf photospheres.  However, these models have difficulty reproducing two outstanding features of this transition: substantial evolution in spectral energy distributions over narrow temperature ($\Delta${\teff} $\approx$ 200--400~K) and luminosity
scales ($\Delta${\logl} $\approx$ 0.3~dex; e.g., \citealt{2004AJ....127.3516G}), and a dramatic increase in 
1~$\mu$m surface brightnesses
(the ``J-band bump''; e.g., \citealt{2003AJ....126..975T}).  Importantly, the latter effect
has been observed in the components of L dwarf/T dwarf binaries, indicating that age, surface gravity or composition variations are not responsible (Figure~\ref{fig_binary}; \citealt{2006ApJS..166..585B, 2006ApJ...647.1393L}.)

These surprising trends suggest a relatively rapid dissolution of 
condensate clouds at the L dwarf/T dwarf transition.
The 1.05 and 1.25~$\mu$m spectral peaks, minima in molecular gas opacity, 
are dominated by condensate opacity in the L dwarfs. A brightening in this region would be expected if 
the condensates were suddenly removed (a deepening of alkali lines and FeH absorption at these flux peaks are also seen across this transition).  But what removes the clouds?
\citet{2001ApJ...556..872A} and \citet{2002ApJ...571L.151B} have proposed a progressive fragmentation of the cloud layer, revealing hot spots similar to those observed on Jupiter and Saturn at 5~$\mu$m; \citet{2004AJ....127.3553K} have proposed a global increase in 
condensate sedimentation efficiency.
Interestingly, atmospheric models by \citet{2006ApJ...640.1063B} predict the formation of a detached
convective region in the photospheres of L dwarfs triggered by dust opacity, which merges with the lower
convective boundary around the L dwarf/T dwarf transition.  What role this transition 
has on cloud structure will require more detailed 3D
atmosphere modeling to simultaneously explore dynamics and cloud formation 
(see contribution by B.\ Freytag).

\section{Clouds and Dynamics in Exoplanet Atmospheres}

\subsection{Empirical Constaints on Cloud Properties}

Like the L dwarfs, exoplanets orbiting close to solar-type stars have equilibrium 
temperatures ({\teq}) spanning phase transitions for several refractory species.
Hence, these objects are also expected to host dusty atmospheres.
Yet observational evidence remains ambiguous.  Transmission spectroscopy has revealed muted atomic and molecular features in some sources (e.g., \citealt{2002ApJ...568..377C, 2008MNRAS.385..109P}) which,
like the {\wat} bands in L dwarf spectra, may be attributable to obscuring condensate (haze) opacity.
In addition, \citet{2007Natur.445..892R} have claimed the detection of silicate emission from HD~20948b, albeit at low significance.
These results stand in contrast to the very low albedo of planets such as HD~20948b, A$_g$$<$0.08 compared to Jupiter's A$_g$$\approx$0.5 \citep{2008ApJ...689.1345R}.  \citet{2008ApJ...682.1277B} have found such albedos  consistent with a condensate-free atmosphere.  On the other hand, 
\citet{2005MNRAS.364..649F} have proposed that the slant angle observations that characterize transit spectroscopy can result in a 1-2 order of magnitude increase in condensate and haze optical depth, allowing a thin condensate layer with little effect on reflectance to absorb strongly in transmission. Since other processes, such as ionization from stellar irradiation or absorption from stratospheric gas ($\S$~3.4) can also
mute photospheric molecular absorption in transmission, the presence and characteristics of condensates in exoplanets remain open questions.

\subsection{Variability}

As with cool dwarfs, variability in exoplanets could support the presence of clouds, particularly in repeated observations of the same longitude (i.e., multiple measurements at a particular phase).
However, while a few studies have suggested the presence of orbit-to-orbit variability in well-studied exoplanets (e.g., \citealt{madhu09}), existing datasets remain too limited in temporal coverage to make definitive claims (long-term warm {\em Spitzer} programs are likely to improve this situation).  Dramatic variations in cloud coverage may also be unlikely, as the typical Rhines lengths and Rossby deformation radii of slowly-rotating, tidally-locked exoplanets are of order the planetary radius \citep{2008ASPC..398..419S}.  
This is the case of circularized planets; eccentric planets, on the other hand, can have dramatic temporal variations in the course of single orbit.  An extreme example is the highly eccentric ($e$ = 0.93) transitting exoplanet HD~80606b, whose brightness temperature was observed to increase by $\sim$700~K over a 6~hr period near periastron \citep{2009Natur.457..562L}.  This system undergoes dramatic changes in irradiation over its 0.3~yr orbit---a factor of over 800 between its 0.03~AU periastron and 0.87~AU apastron---so interorbit variability in its atmospheric properties are almost certain.

\subsection{Dynamics Driven by Irradiation}

\begin{figure}[]
\centering
\resizebox{\hsize}{!}{\includegraphics[clip=true]{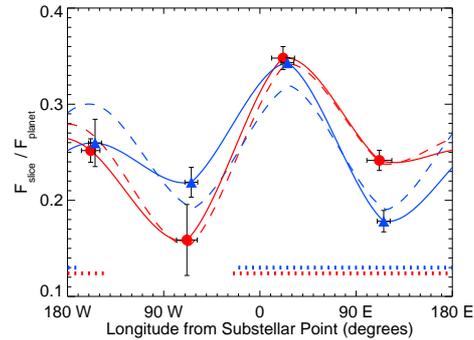}}
\caption{\footnotesize
Longitudinal surface brightness as a percentage of total surface flux for the transitting exoplanet HD~189733b at 8~$\mu$m (blue triangles) and 24~$\mu$m (red circles).  
The brightness peak at both wavelengths is offset from the substellar point by $\sim$20$\degr$, indicating strong horizontal flows.  The relatively modest variation between dayside and nightside flux indicates heat convection around the planet (from \citealt{2009ApJ...690..822K}).
}
\label{fig_exothermal}
\end{figure}

Thermal imaging of some exoplanet atmospheres over significant fractions of their orbits reveal two features indicative of dynamics: thermal peaks offset by 10--30$\degr$ from the substellar longitude and small variations between dayside and nightside brightness temperatures (Figure~\ref{fig_exothermal}; e.g., \citealt{2009ApJ...690..822K}).  Both indicate the presence of winds and jets circulating heat around the planet.  Advanced hydrodynamic models generally confirm these results, although not necessarily the phase amplitudes (e.g., \citealt{2007ApJ...657L.113L, 2009ApJ...699..564S}). Such behavior is not universal; some exoplanets exhibit large day-night flux variations indicative of minimal heat circulation (e.g., \citealt{2006Sci...314..623H}).
In such cases, azimuthal heat gradients will induce azimuthal chemical gradients, likely modulated by the same nonequilibrium chemistry present in brown dwarf
atmospheres (e.g., \citealt{2006ApJ...649.1048C}).

\subsection{Stratospheres and Exospheres}

\begin{figure}[]
\centering
\resizebox{\hsize}{!}{\includegraphics[clip=true, angle=-90]{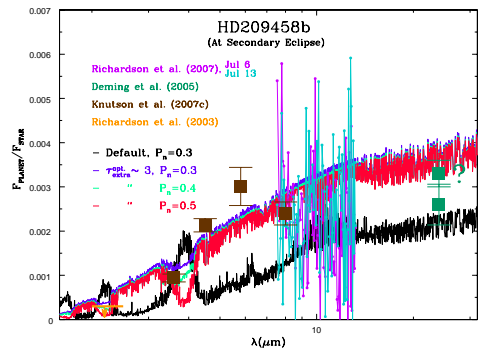}}
\caption{\footnotesize
Planet-to-star flux during secondary transit for HD~209458b.  Mid-infrared
measurements (squares and points with error bars)
are compared to models with (purple, green and red lines) and without (black line) the addition of stratospheric opacity.  Photometry over 4--7~$\mu$m from \citet[labeled 2007c]{2008ApJ...673..526K} 
can only be reproduced with the inclusion of this opacity (from \citealt{2007ApJ...668L.171B}).
}
\label{fig_stratosphere}
\end{figure}

Molecular {\em emission} has been detected during the secondary transit  of a handful of exoplanets
(e.g., \citealt{2008ApJ...673..526K}), indicating the occasional presence of an upper atmosphere temperature inversion, or stratosphere 
(Figure~\ref{fig_stratosphere}; \citealt{2006ApJ...642..495F, 2007ApJ...668L.171B}).
Stratospheres appear to be particularly common in highly irradiated planets (highest {\teq}) and can arise if efficient optical/UV absorbers are present high in the atmosphere. Primary candidates for these absorbers are currently TiO and VO, 
although \citet{2009ApJ...699.1487S} have pointed out that TiO gas requires significant vertical mixing across the condensation cold trap, with eddy coefficients of {\kzz} $\gtrsim$10$^7$~{\cm2s}.  This is somewhat more vigorous than the mixing inferred in brown dwarf atmospheres ($\S$~2.2), butit could be powered by intense stellar irradiation.  The presence of a stratosphere modifies the energy budget, pressure/temperature profile, chemical distribution and emergent spectral flux of an exoplanet, and has been proposed
as a natural division between hot exoplanet types (e.g., \citealt{2008ApJ...678.1419F}).

More controversial is the possible detection of exospheres around exoplanets, atmospheric material blown off by high levels of UV irradiation.   \citet{2003Natur.422..143V} first reported this effect in HD~209485b on the basis of deep Ly$\alpha$ hydrogen absorption during primary eclipse
(or more accurately, a large absorption cross-section).   
Several models have validated this observation (e.g,. \citealt{2003ApJ...598L.121L}), although
\citet{2007ApJ...671L..61B} claim the evaporation detection to be spurious (however, \citealt{2008ApJ...676L..57V} refute this).  The estimated evaporation rate of material off HD~209458b has been estimated to be as high as 0.1~{\mjup}~Gyr$^{-1}$, a value that may be uniquely high given the equivalent mass distributions of exoplanets at close and far separations \citep{2007ApJ...658L..59H}.  Since this process has profound implications on the frequency of Hot Jupiters (which might otherwise be evaporated away), further study is warranted.

\section{Opportunities for Synergy}

The similar properties of cool dwarf and hot exoplanet atmospheres (low temperatures, molecule-rich
atmospheres, presence of clouds and dynamics) but distinct approaches to their
study (e.g, direct observations versus transit detection and phase variation) indicate opportunities for
synergistic efforts in addressing common, outstanding problems.  These include the detailed microphysics of condensate grains (size distribution, composition, spectral properties), the temporal behavior of clouds, the influence of vertical and azimuthal mixing on chemical abundances,
the formation of jets and vorticities, and the opacities of
major molecular species.  Some of these efforts are better addressed in cool dwarf studies, 
due to the ease of detailed, direct observation; others will benefit from the well-characterized
bulk properties (mass and radius) and surface mapping achievable with transiting exoplanets.
As advances are made in detailed, 3D modeling of low-temperature
atmospheres and associated processes, this author encourages 
the use of both cool dwarf and hot exoplanet observations as empirical testbeds.

\begin{acknowledgements}
The author acknowledges helpful comments from J.\ Fortney, N.\ Madhusudhan and D.\ Saumon in the preparation of this review, and thanks E.\ Artigau, A.\ Burrows and H.\ Knutson for electronic versions of their published figures and permission to reproduce them here.
\end{acknowledgements}

\bibliographystyle{aa}


\begin{thebibliography}{57}
\expandafter\ifx\csname natexlab\endcsname\relax\def\natexlab#1{#1}\fi

\bibitem[{{Ackerman} \& {Marley}(2001)}]{2001ApJ...556..872A}
{Ackerman}, A.~S. \& {Marley}, M.~S. 2001, \apj, 556, 872

\bibitem[{{Allard} {et~al.}(2001){Allard}, {Hauschildt}, {Alexander},
  {Tamanai}, \& {Schweitzer}}]{2001ApJ...556..357A}
{Allard}, F., {Hauschildt}, P.~H., {Alexander}, D.~R., {Tamanai}, A., \&
  {Schweitzer}, A. 2001, \apj, 556, 357

\bibitem[{{Artigau} {et~al.}(2009){Artigau}, {Bouchard}, {Doyon}, \&
  {Lafreni{\`e}re}}]{2009ApJ...701.1534A}
{Artigau}, {\'E}., {Bouchard}, S., {Doyon}, R., \& {Lafreni{\`e}re}, D. 2009,
  \apj, 701, 1534

\bibitem[{{Ben-Jaffel}(2007)}]{2007ApJ...671L..61B}
{Ben-Jaffel}, L. 2007, \apjl, 671, L61

\bibitem[{{Burgasser} {et~al.}(2006){Burgasser}, {Kirkpatrick}, {Cruz}, {Reid},
  {Leggett}, {Liebert}, {Burrows}, \& {Brown}}]{2006ApJS..166..585B}
{Burgasser}, A.~J., {Kirkpatrick}, J.~D., {Cruz}, K.~L., {et~al.} 2006, \apjs,
  166, 585

\bibitem[{{Burgasser} {et~al.}(2008){Burgasser}, {Looper}, {Kirkpatrick},
  {Cruz}, \& {Swift}}]{2008ApJ...674..451B}
{Burgasser}, A.~J., {Looper}, D.~L., {Kirkpatrick}, J.~D., {Cruz}, K.~L., \&
  {Swift}, B.~J. 2008, \apj, 674, 451

\bibitem[{{Burgasser} {et~al.}(2002){Burgasser}, {Marley}, {Ackerman},
  {Saumon}, {Lodders}, {Dahn}, {Harris}, \&
  {Kirkpatrick}}]{2002ApJ...571L.151B}
{Burgasser}, A.~J., {Marley}, M.~S., {Ackerman}, A.~S., {et~al.} 2002, \apjl,
  571, L151

\bibitem[{{Burrows} {et~al.}(2007){Burrows}, {Hubeny}, {Budaj}, {Knutson}, \&
  {Charbonneau}}]{2007ApJ...668L.171B}
{Burrows}, A., {Hubeny}, I., {Budaj}, J., {Knutson}, H.~A., \& {Charbonneau},
  D. 2007, \apjl, 668, L171

\bibitem[{{Burrows} {et~al.}(2008){Burrows}, {Ibgui}, \&
  {Hubeny}}]{2008ApJ...682.1277B}
{Burrows}, A., {Ibgui}, L., \& {Hubeny}, I. 2008, \apj, 682, 1277

\bibitem[{{Burrows} {et~al.}(2000){Burrows}, {Marley}, \&
  {Sharp}}]{2000ApJ...531..438B}
{Burrows}, A., {Marley}, M.~S., \& {Sharp}, C.~M. 2000, \apj, 531, 438

\bibitem[{{Burrows} \& {Sharp}(1999)}]{1999ApJ...512..843B}
{Burrows}, A. \& {Sharp}, C.~M. 1999, \apj, 512, 843

\bibitem[{{Burrows} {et~al.}(2006){Burrows}, {Sudarsky}, \&
  {Hubeny}}]{2006ApJ...640.1063B}
{Burrows}, A., {Sudarsky}, D., \& {Hubeny}, I. 2006, \apj, 640, 1063

\bibitem[{{Charbonneau} {et~al.}(2005){Charbonneau}, {Allen}, {Megeath},
  {Torres}, {Alonso}, {Brown}, {Gilliland}, {Latham}, {Mandushev}, {O'Donovan},
  \& {Sozzetti}}]{2005ApJ...626..523C}
{Charbonneau}, D., {Allen}, L.~E., {Megeath}, S.~T., {et~al.} 2005, \apj, 626,
  523

\bibitem[{{Charbonneau} {et~al.}(2002){Charbonneau}, {Brown}, {Noyes}, \&
  {Gilliland}}]{2002ApJ...568..377C}
{Charbonneau}, D., {Brown}, T.~M., {Noyes}, R.~W., \& {Gilliland}, R.~L. 2002,
  \apj, 568, 377

\bibitem[{{Cooper} \& {Showman}(2006)}]{2006ApJ...649.1048C}
{Cooper}, C.~S. \& {Showman}, A.~P. 2006, \apj, 649, 1048

\bibitem[{{Cushing} {et~al.}(2006)}]{2006ApJ...648..614C}
{Cushing}, M.~C. {et~al.} 2006, \apj, 648, 614

\bibitem[{{Deming} {et~al.}(2005){Deming}, {Seager}, {Richardson}, \&
  {Harrington}}]{2005Natur.434..740D}
{Deming}, D., {Seager}, S., {Richardson}, L.~J., \& {Harrington}, J. 2005,
  \nat, 434, 740

\bibitem[{{Fegley} \& {Lodders}(1994)}]{1994Icar..110..117F}
{Fegley}, B.~J. \& {Lodders}, K. 1994, Icarus, 110, 117

\bibitem[{{Fortney}(2005)}]{2005MNRAS.364..649F}
{Fortney}, J.~J. 2005, \mnras, 364, 649

\bibitem[{{Fortney} {et~al.}(2008){Fortney}, {Lodders}, {Marley}, \&
  {Freedman}}]{2008ApJ...678.1419F}
{Fortney}, J.~J., {Lodders}, K., {Marley}, M.~S., \& {Freedman}, R.~S. 2008,
  \apj, 678, 1419

\bibitem[{{Fortney} {et~al.}(2006){Fortney}, {Saumon}, {Marley}, {Lodders}, \&
  {Freedman}}]{2006ApJ...642..495F}
{Fortney}, J.~J., {Saumon}, D., {Marley}, M.~S., {Lodders}, K., \& {Freedman},
  R.~S. 2006, \apj, 642, 495

\bibitem[{{Gelino} \& {Marley}(2000)}]{2000ASPC..212..322G}
{Gelino}, C. \& {Marley}, M. 2000, in Astronomical Society of the Pacific
  Conference Series, Vol. 212, From Giant Planets to Cool Stars, ed. C.~A.
  {Griffith} \& M.~S. {Marley}, 322--+

\bibitem[{{Goldman}(2005)}]{2005AN....326.1059G}
{Goldman}, B. 2005, Astronomische Nachrichten, 326, 1059

\bibitem[{{Golimowski} {et~al.}(2004)}]{2004AJ....127.3516G}
{Golimowski}, D.~A. {et~al.} 2004, \aj, 127, 3516

\bibitem[{{Griffith} \& {Yelle}(1999)}]{1999ApJ...519L..85G}
{Griffith}, C.~A. \& {Yelle}, R.~V. 1999, \apjl, 519, L85

\bibitem[{{Harrington} {et~al.}(2006){Harrington}, {Hansen}, {Luszcz},
  {Seager}, {Deming}, {Menou}, {Cho}, \& {Richardson}}]{2006Sci...314..623H}
{Harrington}, J., {Hansen}, B.~M., {Luszcz}, S.~H., {et~al.} 2006, Science,
  314, 623

\bibitem[{{Helling} {et~al.}(2008){Helling}, {Dehn}, {Woitke}, \&
  {Hauschildt}}]{2008ApJ...675L.105H}
{Helling}, C., {Dehn}, M., {Woitke}, P., \& {Hauschildt}, P.~H. 2008, \apjl,
  675, L105

\bibitem[{{Hubbard} {et~al.}(2007){Hubbard}, {Hattori}, {Burrows}, \&
  {Hubeny}}]{2007ApJ...658L..59H}
{Hubbard}, W.~B., {Hattori}, M.~F., {Burrows}, A., \& {Hubeny}, I. 2007, \apjl,
  658, L59

\bibitem[{{Kalas} {et~al.}(2008){Kalas}, {Graham}, {Chiang}, {Fitzgerald},
  {Clampin}, {Kite}, {Stapelfeldt}, {Marois}, \& {Krist}}]{2008Sci...322.1345K}
{Kalas}, P., {Graham}, J.~R., {Chiang}, E., {et~al.} 2008, Science, 322, 1345

\bibitem[{{Kirkpatrick}(2005)}]{2005ARA&A..43..195K}
{Kirkpatrick}, J.~D. 2005, \araa, 43, 195

\bibitem[{{Knapp} {et~al.}(2004)}]{2004AJ....127.3553K}
{Knapp}, G.~R. {et~al.} 2004, \aj, 127, 3553

\bibitem[{{Knutson} {et~al.}(2008){Knutson}, {Charbonneau}, {Allen}, {Burrows},
  \& {Megeath}}]{2008ApJ...673..526K}
{Knutson}, H.~A., {Charbonneau}, D., {Allen}, L.~E., {Burrows}, A., \&
  {Megeath}, S.~T. 2008, \apj, 673, 526

\bibitem[{{Knutson} {et~al.}(2009){Knutson}, {Charbonneau}, {Cowan}, {Fortney},
  {Showman}, {Agol}, {Henry}, {Everett}, \& {Allen}}]{2009ApJ...690..822K}
{Knutson}, H.~A., {Charbonneau}, D., {Cowan}, N.~B., {et~al.} 2009, \apj, 690,
  822

\bibitem[{{Lagrange} {et~al.}(2009){Lagrange}, {Gratadour}, {Chauvin}, {Fusco},
  {Ehrenreich}, {Mouillet}, {Rousset}, {Rouan}, {Allard}, {Gendron}, {Charton},
  {Mugnier}, {Rabou}, {Montri}, \& {Lacombe}}]{2009A&A...493L..21L}
{Lagrange}, A.-M., {Gratadour}, D., {Chauvin}, G., {et~al.} 2009, \aap, 493,
  L21

\bibitem[{{Lammer} {et~al.}(2003){Lammer}, {Selsis}, {Ribas}, {Guinan},
  {Bauer}, \& {Weiss}}]{2003ApJ...598L.121L}
{Lammer}, H., {Selsis}, F., {Ribas}, I., {et~al.} 2003, \apjl, 598, L121

\bibitem[{{Langton} \& {Laughlin}(2007)}]{2007ApJ...657L.113L}
{Langton}, J. \& {Laughlin}, G. 2007, \apjl, 657, L113

\bibitem[{{Laughlin} {et~al.}(2009){Laughlin}, {Deming}, {Langton}, {Kasen},
  {Vogt}, {Butler}, {Rivera}, \& {Meschiari}}]{2009Natur.457..562L}
{Laughlin}, G., {Deming}, D., {Langton}, J., {et~al.} 2009, \nat, 457, 562

\bibitem[{{Liu} {et~al.}(2006){Liu}, {Leggett}, {Golimowski}, {Chiu}, {Fan},
  {Geballe}, {Schneider}, \& {Brinkmann}}]{2006ApJ...647.1393L}
{Liu}, M.~C., {Leggett}, S.~K., {Golimowski}, D.~A., {et~al.} 2006, \apj, 647,
  1393

\bibitem[{{Lodders}(2002)}]{2002ApJ...577..974L}
{Lodders}, K. 2002, \apj, 577, 974

\bibitem[{{Looper} {et~al.}(2008){Looper}, {Gelino}, {Burgasser}, \&
  {Kirkpatrick}}]{2008ApJ...685.1183L}
{Looper}, D.~L., {Gelino}, C.~R., {Burgasser}, A.~J., \& {Kirkpatrick}, J.~D.
  2008, \apj, 685, 1183

\bibitem[{{Madhusudhan} \& {Seager}(2009)}]{madhu09}
{Madhusudhan}, N. \& {Seager}, S. 2009, ApJ, submitted

\bibitem[{{Marois} {et~al.}(2008){Marois}, {Macintosh}, {Barman}, {Zuckerman},
  {Song}, {Patience}, {Lafreni{\`e}re}, \& {Doyon}}]{2008Sci...322.1348M}
{Marois}, C., {Macintosh}, B., {Barman}, T., {et~al.} 2008, Science, 322, 1348

\bibitem[{{Mohanty} {et~al.}(2002){Mohanty}, {Basri}, {Shu}, {Allard}, \&
  {Chabrier}}]{2002ApJ...571..469M}
{Mohanty}, S., {Basri}, G., {Shu}, F., {Allard}, F., \& {Chabrier}, G. 2002,
  \apj, 571, 469

\bibitem[{{Noll} {et~al.}(1997){Noll}, {Geballe}, \&
  {Marley}}]{1997ApJ...489L..87N}
{Noll}, K.~S., {Geballe}, T.~R., \& {Marley}, M.~S. 1997, \apjl, 489, L87+

\bibitem[{{Pont} {et~al.}(2008){Pont}, {Knutson}, {Gilliland}, {Moutou}, \&
  {Charbonneau}}]{2008MNRAS.385..109P}
{Pont}, F., {Knutson}, H., {Gilliland}, R.~L., {Moutou}, C., \& {Charbonneau},
  D. 2008, \mnras, 385, 109

\bibitem[{{Reiners} \& {Basri}(2008)}]{2008ApJ...684.1390R}
{Reiners}, A. \& {Basri}, G. 2008, \apj, 684, 1390

\bibitem[{{Richardson} {et~al.}(2007){Richardson}, {Deming}, {Horning},
  {Seager}, \& {Harrington}}]{2007Natur.445..892R}
{Richardson}, L.~J., {Deming}, D., {Horning}, K., {Seager}, S., \&
  {Harrington}, J. 2007, \nat, 445, 892

\bibitem[{{Rowe} {et~al.}(2008){Rowe}, {Matthews}, {Seager}, {Miller-Ricci},
  {Sasselov}, {Kuschnig}, {Guenther}, {Moffat}, {Rucinski}, {Walker}, \&
  {Weiss}}]{2008ApJ...689.1345R}
{Rowe}, J.~F., {Matthews}, J.~M., {Seager}, S., {et~al.} 2008, \apj, 689, 1345

\bibitem[{{Saumon} {et~al.}(2006){Saumon}, {Marley}, {Cushing}, {Leggett},
  {Roellig}, {Lodders}, \& {Freedman}}]{2006ApJ...647..552S}
{Saumon}, D., {Marley}, M.~S., {Cushing}, M.~C., {et~al.} 2006, \apj, 647, 552

\bibitem[{{Saumon} {et~al.}(2007){Saumon}, {Marley}, {Leggett}, {Geballe},
  {Stephens}, {Golimowski}, {Cushing}, {Fan}, {Rayner}, {Lodders}, \&
  {Freedman}}]{2007ApJ...656.1136S}
{Saumon}, D., {Marley}, M.~S., {Leggett}, S.~K., {et~al.} 2007, \apj, 656, 1136

\bibitem[{{Showman} {et~al.}(2009){Showman}, {Fortney}, {Lian}, {Marley},
  {Freedman}, {Knutson}, \& {Charbonneau}}]{2009ApJ...699..564S}
{Showman}, A.~P., {Fortney}, J.~J., {Lian}, Y., {et~al.} 2009, \apj, 699, 564

\bibitem[{{Showman} {et~al.}(2008){Showman}, {Menou}, \&
  {Cho}}]{2008ASPC..398..419S}
{Showman}, A.~P., {Menou}, K., \& {Cho}, J.~Y.-K. 2008, in Astronomical Society
  of the Pacific Conference Series, Vol. 398, Astronomical Society of the
  Pacific Conference Series, ed. D.~{Fischer}, F.~A. {Rasio}, S.~E. {Thorsett},
  \& A.~{Wolszczan}, 419--+

\bibitem[{{Spiegel} {et~al.}(2009){Spiegel}, {Silverio}, \&
  {Burrows}}]{2009ApJ...699.1487S}
{Spiegel}, D.~S., {Silverio}, K., \& {Burrows}, A. 2009, \apj, 699, 1487

\bibitem[{{Stephens} {et~al.}(2009){Stephens}, {Leggett}, {Cushing}, {Marley},
  {Saumon}, {Geballe}, {Golimowski}, {Fan}, \& {Noll}}]{2009ApJ...702..154S}
{Stephens}, D.~C., {Leggett}, S.~K., {Cushing}, M.~C., {et~al.} 2009, \apj,
  702, 154

\bibitem[{{Tinney} {et~al.}(2003){Tinney}, {Burgasser}, \&
  {Kirkpatrick}}]{2003AJ....126..975T}
{Tinney}, C.~G., {Burgasser}, A.~J., \& {Kirkpatrick}, J.~D. 2003, \aj, 126,
  975

\bibitem[{{Vidal-Madjar} {et~al.}(2003){Vidal-Madjar}, {Lecavelier des Etangs},
  {D{\'e}sert}, {Ballester}, {Ferlet}, {H{\'e}brard}, \&
  {Mayor}}]{2003Natur.422..143V}
{Vidal-Madjar}, A., {Lecavelier des Etangs}, A., {D{\'e}sert}, J.-M., {et~al.}
  2003, \nat, 422, 143

\bibitem[{{Vidal-Madjar} {et~al.}(2008){Vidal-Madjar}, {Lecavelier des Etangs},
  {D{\'e}sert}, {Ballester}, {Ferlet}, {H{\'e}brard}, \&
  {Mayor}}]{2008ApJ...676L..57V}
{Vidal-Madjar}, A., {Lecavelier des Etangs}, A., {D{\'e}sert}, J.-M., {et~al.}
  2008, \apjl, 676, L57

\end{thebibliography}

\end{document}